\documentstyle[]{article}
\font\cero=cmss10 scaled 1728 \font\uno=cmssbx10 scaled 1200
\begin{document}
\begin{flushleft}
{ \cero Hamiltonian study  for  Chern-Simons  and  Pontryagin theories  }  \\ [3em]
\end{flushleft}
 {\sf { $\star$Alberto Escalante and Leopoldo Carbajal  } }  \\
 {\it   \\ 
 $\star$  Instituto de F{\'i}sica Luis Rivera Terrazas, Benem\'erita Universidad Aut\'onoma de Puebla, (IFUAP). \\
   Apartado postal      J-48 72570 Puebla. Pue., M\'exico, \\ 
    $\star$ LUTh, Observatoire de Paris, Meudon, France,  \\
    Universidad Aut\'onoma Benito Ju\'arez de Oaxaca, Oax., M\'exico,}  \\
  ($\star$  aescalan@sirio.ifuap.buap.mx, alberto.escalante@obspm.fr) \\
 
\noindent{\uno Abstract} \vspace{.5cm}\\
The Hamiltonian analysis  for the Chern-Simons theory and Pontryagin invariant,  which  depends of  a connection valued in the Lie algebra of $SO(3,1)$,  is performed. By applying   a pure Dirac's  method we find for both theories the extended Hamiltonian, the extended action, the constraint algebra,  the gauge transformations and we carry out the counting of degrees of freedom. From the results obtained in the present  analysis, we will conclude that the theories under study have a  closed relation among its   constraints   and defines a topological field theory. In addition,  we extends the configuration space   for the Pontryagin theory and we develop the Hamitonian analysis for this modified version, finding a best description of  the results previously obtained.
\setcounter{equation}{0} \label{c2}
\begin{center}
{\uno I. INTRODUCTION}
\end{center}
\vspace{1em} \
Presently,  the study of topological field theories is a topic of great interest in physics. The importance to study those theories rise  because  shares a  closed  relation with General Relativity. Topological field  theories  are  characterized due to they are  devoid of local physical degrees of freedom,   background independent  and   invariant under diffeomorphisms  \cite{1a}.  Relevant  examples of topological field theories with closed   symmetries  to General Relativity  are  the called $BF$ theories. $BF$  theories  were introduced as generalizations of three dimensional Chern-Simons actions or in other cases,  can also be consider as a zero coupling limit of Yang-Mills theories \cite{1, 2}. We can find in the literature several examples where $BF$ theories comes to be relevant models,  for instance in  alternative    formulations of gravity such as the Pleba\'nski  or  Macdowell-Mansoury.  Pleba\'nski's  formulation consists in to obtain General Relativity  by imposing extra constraints on a $BF$ theory  with the gauge group $SO(3,1)$ or $SO(4)$ \cite{3}. On the other hand, MacDowell-Maunsouri formulation of gravity consists  in  braking down   the $SO$(5) symmetry  of a  $BF$-theory from $SO$(5)  group  to $SO$(4),   to obtain   the Palatini action plus the sum of    second Chern (or Pontryagin class) and Euler topological invariants \cite{4}. Because  those topological classes  have trivial local variations  that do not contribute classically to the dynamics, we  thus obtain essentially general relativity  \cite{5}. \\
Other interesting  theories  reported  in the literature with a closed relation to $BF$ theories,  can be found in the next relation among two functionals   by means  of  \cite{6} \\
\begin{eqnarray}
 \int_{\partial M_{4}} d(A^{IJ} \wedge d A_{IJ} + \frac{2}{3}  A^{IK} \wedge  A_{KL} \wedge A^{L}{_{I}})= \frac{1}{2}\int_{M_{4}} R [A] \wedge R[A],  
 \label{eq1a}
\end{eqnarray}
where  the left hand side can be identified with  the Chern-Simons functional and the right hand side with the Pontryagin class.  Here,     $A^{IJ}$ is  a one-form valued in the Lie algebra of $SO(3,1)$ and  $R^{IJ}$ is the two-form curvature (see below).  As we can see, both   the Chern-Simos and Pontryagin actions are related since the exterior derivative of the former  generates the latter  \cite{6}.  We can observe   in the relation (\ref{eq1a}), that   the  Chern-Simons functional is defined on the boundary of a four dimensional manifold $M_{4}$, while Pontryagin class is defined on $M_{4}$. The study of  the Chern-Simons functional   has been a  topic of several works  because  basically  describes General Relativity  in 3 dimensions  and its  quantization has been worked  out \cite{7}.   Furthermore, by using  the Chern-Simons functional we can construct a wave function  that  corresponds to an   exact state of the Schrodinger equation for Yang-Mills theory  in four dimensions \cite{8a}. In addition,  we can find a recent  work  where  the Chern-Simons state describes a topological state with unbroken diffeomorphism  invariance in Yang-Mills  and General Relativity  \cite{8}. In the loop quantum gravity  context,   that state is called the Kodama state  and has been studied in interesting works by Smolin,   arguing  that the Kodama state at least for the Sitter spacetime, loop quantum gravity does have a good low energy limit  \cite{9}.  On the other hand,  the Pontryagin invariant is another interesting topological field theory   \cite{11, 11b} and  has  been topic of study  in  recently works because  is  expected to be related to physical observables,  as for instance in the case of anomalies \cite{11a, 12, 13, 14, 15, 16}.\\
With  these antecedents in mind,  the purpose of this paper  is to report a pure Dirac's method of constrained systems applied for the actions involved in the relation (\ref{eq1a}),   which is absent in the literature. There  are several reasons to develop this work.  The first one,  we will perform    a pure Dirac analysis which means that we will  work  with the full configuration space and therefore with the full phase space. In other words,   we will consider  all   the  set of one-forms $"A^{IJ}"$ that defines our theories as dynamical ones.  Thus,  with the present study we  will be able  to  known the relation among the actions   at  Lagrangian  level  as well as  at  Hamiltonian level. With the analysis at hand,  we will can   identify  the relevant symmetries for both  theories   for example,   the constraints, the extended action, the extended Hamiltonian, the constrained algebra and  the gauge transformations. In particular,  with all the constraints  classified as first or second class, we will be able to  carry out the    counting of the physical degrees of freedom. The second one, with the present analysis we wish to report a complete study of the  relation among the constraints  that there exists in  the Chern-Simons  theory  and the constraints for the Pontryagin invariant. We can find in recent  results  by developing   Dirac's  quantization or covariant canonical  program for the Pontryagin invariant,  that the Chern-Simons wave function  represents a quantum state for theory \cite{11, 11b}, but the analysis reported in \cite{11, 11b}  has been developed  on a smaller  phase space  and  the full constraints program was not performed. Therefore,  the results of this paper intend  to extend and complete these results by performing a  complete Dirac  analysis where we shall  work with the full   phase  space   reproducing in particular the results found in \cite{11}.  It is important to remark,  that usually  can be found in the literature  the Dirac's analysis applied to  several theories  \cite{17} ,   but  generally    the way to  perform the study is  on a smaller  phase space context,  this means that  the    dynamical variables  are  considered  as  those variables  that occurs  with  temporal derivative in the Lagrangian density \cite{11c}. However,  is not common to find a pure Dirac's method (working with the complete   phase  space)  for field theories \cite{18a}. The principal  reasons for studying  the  Hamiltonian formalism  under a   smaller phase space context and not carried out  on the complete phase space,  is  because  the separation of   the constraints into  first or second class is not easy to carry out. In this manner,  in the  literature  we find that the people  prefer to  work on a smaller phase space context because  generally there are present  only  first class constraints and  is common to avoid   the difficult  part of  the separation among   the  constraints. The price that we pay  by work on a smaller phase space context   is that we can not neither know the complete  form of the constraints,    the complete form of gauge transformations  defined on the full space phase nor the complete algebra among the constraints for the theory under study. Of course, by working with the full configuration  space we can reproduce  the results obtained by working on a smaller configuration  space.  \\
In this manner, because of  the previous explanation  in this work we will perform a pure Dirac method for  the theories expressed in (\ref{eq1a}), obtaining as relevant results  the    complete identification   of its  symmetries.  All this  part  will be clarified  along  the present work .  \\
The paper  is organized as follows: In the Section II we will perform by  using a pure Dirac method the Hamiltonian analysis for the Chern-Simons action. We will identify the full constraints for the theory, the extended action, the extended Hamiltonian, the gauge transformations  and we will carry out the counting of degrees of freedom,  concluding that this theory is devoid of degrees of freedom as  is expected.  In particular, we will show the way  to  identify  the  first and second class constraints and  then   compute the algebra among them.  In Section II we will develop the Hamiltonian analysis for the Pontryagin invariant expressed as in (\ref{eq1a}). We will find the extended action, the extended Hamiltonian, the full constraints program, the gauge transformations  and the counting of degrees of freedom,  allow us to conclude that the theory is a topological field theory too. As important part of this section,  we will find that contrary to Chern-Simons theory  the Pontryagin invariant presents only a set of first class constraints.  In Section III we will extend the configuration  space for the Pontryagin invariant and we will perform the Hamiltonian analysis  for this modified theory. As important result that we will find in this section is that we will have a best description than   the results obtained  above, but the price to pay for this description is that contrary to Section II, now we will have the presence of first and second class constraints. In particular, we will reproduce the results found previously considering the second class constraints as strong equations.   
\newline
\newline
\noindent \textbf{I. Hamiltonian dynamics for  the Chern-Simons term  }\\[1ex]
In this section, we will  perform the Hamiltonian dynamics   for the Chern-Simons term which  will   be    expressed  by \cite{11}
\begin{equation}
S[A]_{C-S}= \frac{\alpha}{2} \int_M A^{IJ} \wedge d A_{IJ} + \frac{2}{3}  A^{IK} \wedge  A_{KL} \wedge A^{L}{_{I}}, 
\label{eq2}
\end{equation}
here, $A^{IJ}=A{_{\mu}}^{IJ}dx^{\mu}$ is the Lorentz  connection valued in the  Lie algebra of $SO(3,1)$, $\mu, \nu=0,1,2$ are spacetime indices, $x^\mu$  are the coordinates that label the points for the 3-dimensional Minkowski manifold $M$ and  $I, J= 0,1..,3$ are internal indices that can be raised and lowered by the internal Lorentzian   metric  $\eta_{IJ}= (-1,1,1,1)$.\\ 
We start  computing the Euler-Lagrange equations  obtained from the variation of the  action (\ref{eq2}), which  are given by 
\begin{equation}
\epsilon^{\alpha \beta \mu} F_{\beta \mu IJ}=0, 
\label{eq3}
\end{equation}
where, $F_{\beta \mu IJ}=\partial_\beta A_{\mu IJ}-\partial_\mu A_{\beta IJ}+ A_{\mu I K} A_\beta{^{K}}_{J}-A_{\beta I K} A_\mu{^{K}}_{J} $.  The equations of motion (\ref{eq3}) whose solutions corresponds to the space of  flat connections,   will be useful to identify  the gauge transformations for the theory,  work that  will be developed below.   \\
Now, we will consider that the manifold $M$ has a topology $\Sigma \times R$, where $\Sigma$ corresponds to a Cauchy's surface. By using this fact,  we  perform   the 2+1 decomposition in the action (\ref{eq2})   obtaining  
\begin{equation}
S[A]_{C-S} = \int_M \left[  \frac{\alpha}{2} \epsilon^{0ab} A_{0}{^{IJ}} F_{abIJ} + \frac{\alpha}{2} \epsilon^{0ab} A_{b}{^{IJ}} \dot{A}_{a IJ} \right] dx^3, 
\label{eq4}
\end{equation}
where $F_{abIJ }= \partial_a A_{bIJ} -\partial_b A_{aIJ} + A_{aI}{^{L}}A_{bLJ}-A_{bI}{^{L}}A_{aLJ} $,  with  $a,b,c=1,2$.  From (\ref{eq4}) we can identify the next Lagrangan density for the Chern-Simons theory
\begin{equation}
{\mathcal{L}} = \frac{\alpha}{2} \epsilon^{0ab} A_{0}{^{IJ}} F_{abIJ } + \frac{\alpha}{2} \epsilon^{0ab} A_{b}{^{IJ}} \dot{A}_{a IJ}.
\label{eq5}
\end{equation}
At this step,  it is common to find  in the  literature  that   the Hamiltonian analysis for the action (\ref{eq4})  is  performed   on a smaller phase space context.  This means that the dynamical variables  are  considered    those one-forms $A^{IJ}$'s  that occurs   in the action with  temporal derivative; in others words,   the follow 12 one-forms  $\rightarrow$$A_{a IJ}$  are identified as   dynamical variables for the action (\ref{eq4}),     and   the rest  6 one-forms  $\rightarrow$$A_{0}{^{IJ}}$ are  identified as  Lagrange multipliers.   Nevertheless, in this work we will develop a pure Dirac method which means that we will consider  our dynamical variables   the set of  $A^{IJ}$'s$=(A_{a IJ}, A_{0}{^{IJ}})$ that defines our theory.  Therefore,   a pure Dirac's method calls for the definition of the momenta $(\Pi^{\alpha}{_{IJ}})$ canonically conjugate to $(A_{\alpha} {^{IJ}})$ 
\begin{equation}
\Pi^{\alpha}{_{IJ}}= \frac{\delta {\mathcal{L}} }{ \delta \dot{A}_{\alpha} {^{IJ}} } .
\label{eq6}
\end{equation}
The matrix elements of the Hessian 
\begin{equation}
\frac{\partial^2{\mathcal{L}} }{\partial \partial_\mu (A_{\alpha} {^{IJ}} ) \partial\partial_\mu (A_{\beta} {^{IJ}} ) } ,
\label{eq7}
\end{equation}
are identically zero, the rank of the Hessian is zero, thus, we expect 18 primary constraints. From the definition of the momenta (\ref{eq6}) we identify the next 18 primary constraints 
\begin{eqnarray}
\phi^{0}{_{IJ}}&:=& \Pi^0{_{IJ}} \approx 0 ,\nonumber \\
\phi^{a}{_{IJ}}&:=& \Pi^a{_{IJ}} - \frac{\alpha}{2}\epsilon^{0ab} A_{b IJ} \approx 0.
\label{eq8}
\end{eqnarray}
We can observe that  by working   on a smaller phase space context  (the dimension of  this smaller space is  24, 12$\rightarrow$$\dot{A}_{a IJ}$  and its   respective momenta)   the first constraint related with $\phi^{0}{_{IJ}}$  is not  taken in to account. However,  the purpose of this paper is  to   work with the full phase space and therefore with the 18 primary constraints (\ref{eq8}).  May be for the lector is not relevant this part, but  once  finished the analysis  for the Chern-Simons and Pontryagin theory,  we will be able to  appreciate the advantage to perform a pure Dirac method,  because we will can  identify   the extended action, the  extended Hamiltonian, the complete form of the constrains and the algebra among them. The correct identification of the constrains is very important because can be used  to  carry out the counting of the physical degrees of freedom. On the other hand,  constraints are  the guideline     to make the best progress for  the quantization of the theory. We need to remember that the   quantization  scheme   for  theories as Maxwell or Yang-Mills can not be directly applied to  theories with the symmetry  of invariance under diffeomorphisms  (as for  instance  topological field theories)   because we can lose  relevant physical information \cite{11}. \\  
By following with the method,  the canonical Hamiltonian for the Chern-Simons system is given by 
\begin{equation}
H_{c}=\int  dx^2 \left[ \dot{A}_{\alpha}{^{IJ}}\Pi{^{\alpha}}_{IJ}- {\mathcal{L}}\right]=  - \int  dx^2 \left[ \frac{\alpha}{2} A_{0}{^{IJ}} \epsilon^{0ab} F_{abIJ} \right].
\label{eq9}
\end{equation}
In this manner, the primary Hamiltonian will be constructed by adding the primary constraints (\ref{eq8}) to (\ref{eq9}),  this is 
\begin{equation}
H_P= H_c + \int dx^2 \left[  \lambda^{IJ}{_{0}} \phi^0{_{IJ}} + \lambda^{IJ}{_{a}} \phi^a{_{IJ}}  \right],
\label{eq10} 
\end{equation}
where $ \lambda^{IJ}{_{0}}$ and $\lambda^ {IJ}{_{a}} $  are Lagrange multipliers enforcing the constraints. The non-vanishing fundamental brackets for our theory are given by 
\begin{equation}
\{ A_{\alpha} {^{IJ}}(x),\Pi^{\beta}{_{KL}}(y) \} = \frac{1}{2} \delta {^ \beta} _\alpha \left( \delta^I {_K} \delta^J{_L} - \delta^I{ _L} \delta^J{_K } \right) \delta^2(x-y).
\label{eq11}
\end{equation}
Now,  we compute the 18 $\times$ 18 matrix whose entries are the Poisson brackets among the constraints (\ref{eq8})  
\begin{eqnarray}
\{\phi^0{_{IJ}} (x),\phi^0{_{KL}} (y) \} &=& 0, \nonumber \\
 \{\phi^0{_{IJ}} (x),\phi^a{_{KL}} (y) \} &=& 0, \nonumber \\
\{\phi^a{_{IJ}} (x),\phi^a{_{KL}} (y) \} &=& 0, \nonumber \\ 
\{\phi^a{_{IJ}} (x),\phi^{b}{_{KL}} (y) \} &=&  \frac{\alpha}{2}\epsilon^{0ab}\left(\eta_{IL}\eta_{JK}- \eta_{IK}\eta_{JL} \right ) \delta^2(x-y),
\label{eq12}
\end{eqnarray}
we can appreciate that this matrix has rank=12  and 6 linearly independent null-vectors. By using the 6 null-vectors and  consistency conditions we arrive  to the next 6 secondary constraints 
\begin{equation}
\dot{\phi}^0{_{IJ}}= \{\phi^0{_{IJ}} (x), {H}_{P} \} \approx 0 \quad \Rightarrow \quad \psi_{IJ}:= \frac{\alpha}{2}\epsilon^{0ab}F_{abIJ}  \approx 0. 
\label{eq13}
\end{equation}
Consistency requires that their conservation in the time vanish as well. For this theory there no, third constraints.  Now, we need to identify from  the primary and secondary constrains which ones corresponds to  first and second class. For this aim,  we need to calculate the rank and the null-vectors of the  24$\times$ 24 matrix whose entries will be the Poisson brackets among  primary and secondary constraints, this is 
\begin{eqnarray}
\{\phi^0{_{IJ}} (x),\phi^0{_{KL}} (y) \} &=& 0, \nonumber \\
 \{\phi^0{_{IJ}} (x),\phi^a{_{KL}} (y) \} &=& 0, \nonumber \\
 \{\phi^0{_{IJ}} (x),\Psi{_{KL}} (y) \} &=& 0, \nonumber \\
\{\phi^a{_{IJ}} (x),\phi^0{_{KL}} (y) \} &=& 0, \nonumber \\ 
\{\phi^a{_{IJ}} (x),\phi^b{_{KL}} (y) \} &=&  \frac{\alpha}{2}\epsilon^{0ab}\left(\eta_{IL}\eta_{JK}- \eta_{IK}\eta_{JL} \right ) \delta^2(x-y),
\nonumber\\
 \{\phi^a{_{IJ}} (x),\Psi{_{KL}} (y) \} &=&  \frac{\alpha}{2}\epsilon^{0ab}\Big\{\left(\eta_{KI}\eta_{LJ}-\eta_{KJ}\eta_{LI} \right) \partial_b \delta^2(x-y) + (\eta_{KJ} A_{bIL} - \eta_{KI}A_{bJL} )\delta^2(x-y) \nonumber \\  
 &-& (\eta_{LI} A_{bKJ} -\eta_{LJ}A_{bKI})\delta^2(x-y) \Big\},  
 \label{eq14}
\end{eqnarray}
this matrix has rank=12 and 12 null-vectors. From the null vectors we can   identify the next  12  first class constraints 
\begin{eqnarray}
\gamma^0{_{IJ}} &:=& \phi^0{_{IJ}}  \approx 0 , \nonumber \\
\gamma{_{IJ}} &:=& \Psi{_{IJ}} + D_a\phi^a{_{IJ}}   \approx 0, 
\label{eq15}
\end{eqnarray}
Here, we can identify that $\gamma{_{IJ}}$ takes the role of  Gauss constraint for the Chern-Simons theory.  On the other hand, the  rank  yields to  identify the next 12 second class constraints 
\begin{equation}
\chi^a{_{IJ}} := \phi^a{_{IJ}}  \approx 0.
\label{eq16}
\end{equation}
The correct identification of  first and  second class constraints allow us to carry out the counting of degrees of freedom in the next form; we have 36 canonical variables  $(A_{\alpha} {^{IJ}},\Pi^{\alpha}{_{IJ}} )$, 12 first class constraints $(\gamma^0{_{IJ}}, \gamma{_{IJ}})$  and 12 second class constraints $(\chi^a{_{IJ}} )$ which yields to conclude that Chern-Simons theory is devoid of degrees of  freedom. Therefore,  defines a topological field theory. \\
To compute the  algebra of  constraints is convenient to smear them  
\begin{eqnarray}
\phi_1 &:=& \gamma^0{_{IJ}} \left[ A \right]= \int dx^2  A^{IJ}  \Pi^0{_{IJ}}, \nonumber \\
\phi_2  &:=& \gamma{_{IJ}}\left[ B \right]= \int dx^2 B{^{IJ}} \left[  \Psi{_{IJ}} + D_a\phi^a{_{IJ}} \right], \nonumber \\
\phi_3  &:=&\chi^a{_{IJ}} \left[ C \right]= \int dx^2 C_a{^{IJ}} \left[  \Pi^a{_{IJ}} - \frac{\alpha}{2}\epsilon^{0ab} A_{b IJ} \right],
\end{eqnarray}
In this manner,  the algebra is 
\begin{eqnarray}
\Big \{ \phi_1 \left[ B^{IJ} \right],\phi_1 \left[ C^{KL} \right] \Big \}&=&  0 , \nonumber \\ 
\Big \{ \phi_1 \left[ B_{IJ} \right],\phi_2 \left[ G^{{IJ}} \right]  \Big \}&=&  0, \nonumber \\
\Big \{ \phi_1 \left[ B{_{IJ}} \right],\phi_3 \left[ G_a{^{KL}} \right] \Big  \}& =& 0, \nonumber \\
\Big  \{ \phi_2 \left[ B^{{IJ}} \right],\phi_2 \left[ G{^{KL}} \right] \Big  \}& =& \int dx^2\left[ B^I{_K}G^{KJ}-B^J{_K}G^{KI}\right]  \gamma{_{IJ}} \approx0 , \nonumber \\
\Big   \{ \phi_2 \left[ B^{{IJ}} \right],\phi_3 \left[ C{_a}{^{KL}} \right] \Big  \}& =& \int dx^2\left[ B^I{_K}C{_a}^{KJ}-B^J{_K}C{_a}^{KI}\right]  \chi^a{_{IJ}} \approx0, \nonumber \\
    \Big \{ \phi_3 \left[ C{_a}^{{IJ}} \right],\phi_3 \left[ G{_b}{^{KL}} \right] \Big  \}& =& - \frac{\alpha}{2}\int dx^2 \epsilon^{0ab}\left[ C_{a IJ}G{_b}^{IJ}-C_{b IJ}G{_a}^{IJ}\right],
    \label{eq18}
\end{eqnarray}
where we can see that the algebra is closed.\\
By identifying  the  first class and second class constraints,  we can find the extended action given by 
\begin{eqnarray}
S_E[A_\alpha{^{IJ}}, \Pi^{\alpha}{_{IJ}}, \lambda_0{^{IJ}}, \lambda^{IJ}] 
& = & \int dx^{3} \Big[ \Pi^{0}{_{IJ}} \dot{A}_0{^{IJ}} + \Pi^{a}{_{IJ}} \dot{A}_a{^{IJ}} - H - \lambda_0{^{IJ}} \gamma^{0}{_{IJ}} - \lambda^{IJ}\gamma_{IJ} \nonumber \\  &-& \upsilon_a{^{IJ}} \chi^{a}{_{IJ}} \Big]
\label{eq19}
\end{eqnarray}
where $H=-A_{0}{^{IJ}} \gamma_{IJ}$,  which is proportional to Gauss first class constraint, and 
 \begin{equation}
H_E = H + \lambda_0{^{IJ}} \gamma^{0}{_{IJ}} + \lambda^{IJ}\gamma_{IJ}, 
\label{eq20}
\end{equation}
being the extended Hamiltonian,  which a linear combination of  first class constraints. As we known,  the equations of motion obtained from the extended Hamiltonian are not equivalent  with  Euler-Lagrange equations, but the difference is unphysical \cite{6}.\\
Now,  we shall compute the equations of motion obtained from the extended action (\ref{eq19}),  which are given by 
\begin{eqnarray}
\delta A_0{^{IJ}} : \dot{\Pi}^{0}{_{IJ}} & = &  \gamma_{IJ}, \nonumber  \\
\delta \Pi^{0}{_{IJ}} : \dot{A}_0{^{IJ}} & = & \lambda_0{^{IJ}}, \nonumber \\
\delta \Pi^{a}{_{IJ}} : \dot{A}_a{^{IJ}} & = & D_a(A_0{^{IJ}}-\lambda^{IJ}) + \upsilon_a{^{IJ}}, \nonumber \\
\delta A_a{^{IJ}} : \dot{\Pi}^{a}{_{IJ}} & = & \frac{1}{2}\epsilon^{0ba}\upsilon_{bIJ} - \frac{1}{2}\epsilon^{0ba}\partial_b(A_{0IJ} - \lambda_{IJ}) - (A_{0I}{^{L}} - \lambda_I{^{L}})\Pi^{a}{_{LJ}}, \nonumber \\
& + &  (A_{0J}{^{L}} - \lambda_J{^{L}})\Pi^{a}{_{LI}}, \nonumber \\
\delta \lambda_0{^{IJ}} : \gamma^{0}{_{IJ}} & = & 0, \nonumber \\
\delta \lambda^{IJ} : \gamma_{IJ} & = & 0,  \nonumber \\
\delta \upsilon_a{^{IJ}} : \chi^{a}{_{IJ}} & = & 0. 
 \label{eq21}
\end{eqnarray}
\newline
\noindent \textbf{I.I Gauge generator }\\[1ex]
 One of the most important symmetries that we can study by using the Hamiltonian method,   are the gauge transformations. Gauge transformations are an important  symmetry,  because they  can  help us to identify  physical observables \cite{17}.  Thus,  we need   to know explicitly the  gauge transformations  for our theory. For this aim,  we will apply the Castellani's algorithm \cite{ 17} to construct the  gauge  generator.  We define the generator of gauge transformations as 
\begin{equation}
G= \int dx^2 \left(D_0 \varepsilon{_{0}}^{IJ} \gamma^{0}{_{IJ}} +   \varepsilon^{IJ} \gamma_{IJ} \right), 
\end{equation}
thus, we can identify the next gauge transformations on the phase space 
\begin{eqnarray}
\delta A_0{^{IJ}} & = & D_0 \varepsilon_0{^{IJ}},  \\
\delta A_b{^{IJ}} & = & - D_b \varepsilon^{IJ},  \\
\delta \Pi^{0}{_{IJ}} & = & - \varepsilon_I{^{L}} \Pi^{0}{_{LJ}} + \varepsilon_J{^{L}} \Pi^{0}{_{LI}}, \\
\delta \Pi^{a}{_{IJ}} & = & \frac{1}{2}\epsilon^{0ba} \partial_b \varepsilon_{IJ} + \Pi^{a}{_J}{^{L}} \varepsilon_{LI} - \Pi^{a}{_I}{^{L}} \varepsilon_{LJ} .
\label{eq22}
\end{eqnarray}
On the other hand, we know that Chern-Simons theory shares the symmetries of general relativity \cite{7} namely,  background independence and diffeomorphisms.  So,  we can formulate the next question; what about the diffeomorphisms in our theory?. Apparently diffeomorphisms are not an internal symmetry, but that is not true at all because we can take $\varepsilon_0{^{IJ}}= - \varepsilon^{IJ}$  and introducing   the new gauge parameters  as  \cite{6}
\begin{equation}
\varepsilon^{IJ}= -\xi^\alpha A_\alpha{^{IJ}},  
\end{equation}
 we obtain  
\begin{equation}
A_{\mu}{^{IJ}} \rightarrow A_{\mu}{^{IJ}} + {\mathcal{L}}_\xi  A_{\mu}{^{IJ}} + \xi^\alpha F_{\mu \alpha}{^{IJ}}.
\label{eq23}
\end{equation}
Therefore, diffeomorphisms corresponds to an internal symmetry of the theory. \\
As conclusion of this part, we have performed the Hamiltonian analysis for the Chern-Simons theory by working with the complete   configuration   space. With  the present analysis,  we have obtained the extended action, the extended Hamiltonian, the full  constraints program  and the algebra among them. With all these results at hand,  we could  confirm  that Cher-Simons  action  is a topological field theory  and shares symmetries with   General Relativity  as for instance,   diffeormorphisms as gauge transformations. It is important to note  that this theory presents  a set of first and second class constraints. However,  we will see in the next section that   Pontryagin theory presents only a set  of  first class constraints   and reducibility conditions among  them. This fact will be important because  Pontryagin theory is defined in four dimensions. Nevertheless,   we do not lose the symmetries of  Chern-Simons theory which is defined in three dimensions. This fact will be clarified below. 
\newline
\newline
\newline
\noindent \textbf{II. Hamiltonian dynamics for  the Pontryagin invariant }\\[1ex]
In this section, we will perform a pure  Hamiltonian dynamics for the Pontryagin invariant \cite{11, 18} which is absent in the literature. \\
We start with the Pontryagin action expressed as the action (\ref{eq1a})
\begin{equation}
S[A]= \alpha \int_M R^{IJ}[A]\wedge R_{IJ}[A], 
\label{eq26}
\end{equation}
where $R^{IJ}[A]= \frac{1}{2}  R{_{\mu \nu}}^{IJ} dx^\mu \wedge dx^\nu$
is the curvature  of the $SO(3,1)$ 1-form connection $A_{\nu}{^ {IJ}}$ with  $R{_{\mu \nu}}^{IJ}= \partial_\mu A_{\nu}{^ {IJ}} - \partial \nu A_{\mu}{^ {IJ}} + A_{\mu}{^ {IK}}A_{\nu}{_K{^J}}-A_{\nu}{^ {IK}}A_{\mu}{_K{^J}}$. Here,  $\mu, \nu=0,1,..,3$ are spacetime indices, $x^\mu$  are the coordinates that label the points for the 4-dimensional Minkowski manifold $M$ and  $I, J= 0,1..,3$ are internal indices that can be raised and lowered by the internal Lorentzian   metric  $\eta_{IJ}= (-1,1,1,1)$.\\
The equations of motion obtained from the variation of the action (\ref{eq26}) are given by 
\begin{equation}
DR=0,
\label{eq26a}
\end{equation}
where we can see that these equations corresponds to Bianchi identities. \\ 
By performing the $3+1$ decomposition of (\ref{eq26}) we find 
\begin{equation}
S[A]= \alpha  \int dt \int dx^3 \eta^{abc} R_{bcIJ }\left(\dot{A}_a{^{IJ}}- D_a A_0{^{IJ}}  \right), 
\label{eq27}
\end{equation}
here, $a,b,c=1,..,3$,  $R_{abIJ }= \partial_a A_{bIJ} -\partial_b A_{aIJ} + A_{aI}{^{L}}A_{bLJ}-A_{bI}{^{L}}A_{aLJ} $ and $D_a A_b{^{IJ}}= \partial_a A_b^{IJ} + A_a^{IK}A_{bK}{^{J}} + A_a{^{JK}}A_b{^{I}}_K$. \\ 
From (\ref{eq18}) we can identify the next  Lagrangian density 
\begin{equation}
{\mathcal{L}} =  \alpha \eta^{abc} R_{bcIJ }\left(\dot{A}_a{^{IJ}}- D_a A_0{^{IJ}}  \right). 
\label{eq28}
\end{equation}
Just as in  the last section, a pure Dirac's method calls for  the definition of the momenta $(\Pi^{\alpha}{_{IJ}})$ canonically conjugate to $(A_{\alpha} {^{IJ}})$ 
\begin{equation}
\Pi^{\alpha}{_{IJ}}= \frac{\delta {\mathcal{L}} }{ \delta \dot{A}_{\alpha} {^{IJ}} } .
\label{eq29}
\end{equation}
The matrix elements of the Hessian 
\begin{equation}
\frac{\partial^2{\mathcal{L}} }{\partial \partial_\mu (A_{\alpha} {^{IJ}} ) \partial\partial_\mu (A_{\beta} {^{IJ}} ) } ,
\label{eq30}
\end{equation}
are identically zero, the rank of the Hessian is zero, thus, we expect 24 primary constraints. From the definition of the momenta (\ref{eq29}) we identify the next 24 primary constraints 
\begin{eqnarray}
\phi^0{_{IJ}}&:=& \Pi^0{_{IJ}} \approx 0 ,\nonumber \\
\phi^a{_{IJ}}&:=& \Pi^a{_{IJ}} - \alpha  \eta^{abc} R_{bcIJ }  \approx 0.
\label{eq31}
\end{eqnarray}
Neglecting terms  on the frontier, the canonical Hamiltonian for the second Chern class   system is given by 
\begin{equation}
H_{c}= - \int  dx^3 A_0 {^{IJ}}D_a \Pi^a{_{IJ}}.
\label{eq32}
\end{equation}
In this manner,  we add the primary constraints to identify the primary Hamiltonian, given by 
\begin{equation}
H_P= H_c + \int dx^3 \left[  \lambda^{IJ}{_{0}} \phi^0{_{IJ}} + \lambda^{IJ}{_{a}} \phi^a{_{IJ}}  \right],
\label{eq33} 
\end{equation}
where $ \lambda^{IJ}{_{0}}$ and $\lambda^ {IJ}{_{a}} $  are Lagrange multipliers enforcing the constraints. The non-vanishing fundamental brackets are 
\begin{equation}
\{ A_{\alpha} {^{IJ}}(x),\Pi^{\beta}{_{KL}}(y) \} = \frac{1}{2} \delta {^ \beta} _\alpha \left( \delta^I {_K} \delta^J{_L} - \delta^I{ _L} \delta^J{_K } \right) \delta^3(x-y).
\label{eq34}
\end{equation}
Now we compute the 24 $\times$ 24 matrix whose entries are the Poisson brackets among the constraints (\ref{eq31})  
\begin{eqnarray}
\{\phi^0{_{IJ}} (x),\phi^0{_{KL}} (y) \} &=& 0, \nonumber \\
 \{\phi^0{_{IJ}} (x),\phi^a{_{KL}} (y) \} &=& 0, \nonumber \\
\{\phi^a{_{IJ}} (x),\phi^0{_{KL}} (y) \} &=& 0, \nonumber \\ 
\{\phi^a{_{IJ}} (x),\phi^b{_{KL}} (y) \} &=& 0,
\label{eq35}
\end{eqnarray}
we can observe that this part is quite different respect to Chern-simons theory because the entries of the matrix (\ref{eq35}) are all equal to zero.  This means that we can  determine  all  the values of the Lagrange multipliers al most weakly \cite{ 17}. However,  consistency  allow us to identify  the next  6  reducibility conditions 
\begin{equation}
\dot{\phi}^0{_{IJ}}= \{\phi^0{_{IJ}} (x), {H}_{P} \} \approx 0 \quad \Rightarrow \quad \Psi_{IJ}:= D_a \Pi^a{_{IJ}}  \approx 0, 
\label{eq36}
\end{equation}
where can be identified as the Gauss constraint for the theory.  In addition, for this theory there no, third constraints. \\
To compute the  algebra among  the constraints  is convenient  rewrite them  as 
\begin{eqnarray}
\phi_1 &:=& \gamma^0{_{IJ}} \left[ A \right]= \int dx^3  A^{IJ}  \Pi^0{_{IJ}}, \nonumber \\
\phi_2  &:=& \gamma{_{IJ}}\left[ B \right]= \int dx^3 B{^{IJ}} \left[ D_a \Pi^a{_{IJ}}  \right], \nonumber \\
\phi_3  &:=&\gamma^a{_{IJ}} \left[ C \right]= \int dx^3 C_a{^{IJ}} \left[   \Pi^a{_{IJ}} - \alpha  \eta^{abc} R_{bcIJ }  \right],
\label{eq37}
\end{eqnarray}
In this manner,  the algebra is 
\begin{eqnarray}
\Big \{ \phi_1 \left[ B^{IJ} \right],\phi_1 \left[ C^{KL} \right] \Big \}&=&  0 , \nonumber \\ 
\Big \{ \phi_1 \left[ B{IJ} \right],\phi_2 \left[ G^{{IJ}} \right]  \Big \}&=&  0, \nonumber \\
\Big \{ \phi_1 \left[ B{_{IJ}} \right],\phi_3 \left[ G_a{^{KL}} \right] \Big  \}& =& 0, \nonumber \\
\Big  \{ \phi_2 \left[ B^{{IJ}} \right],\phi_2 \left[ G{^{KL}} \right] \Big  \}& =& \int dx^3\left[ B^I{_K}G^{KJ}-B^J{_K}G^{KI}\right]  \gamma{_{IJ}} \approx0 , \nonumber \\
\Big   \{ \phi_2 \left[ B^{{IJ}} \right],\phi_3 \left[ C{_a}{^{KL}} \right] \Big \}& =& \int dx^3\left[ B^I{_K}C{_a}^{KJ}-B^J{_K}C{_a}^{KI}\right]  \gamma^a{_{IJ}} \approx0, \nonumber \\
\Big     \{ \phi_3 \left[ C{_a}^{{IJ}} \right],\phi_3 \left[ G{_b}{^{KL}} \right] \Big  \}& =&0,
\end{eqnarray}
where we can see that the constraints form a first class set. The identification of the constraints allow us carry out the counting of degrees of freedom as follows:  We have 
48 canonical variables $(A_{\alpha} {^{IJ}},\Pi^{\alpha}{_{IJ}} )$ and 30   first class constraints  $(\gamma^0{_{IJ}}, \gamma^a{_{IJ}}, \gamma_{IJ}$). However,  Bianchi's identities $DR=0$ implies  6 reducibility conditions among the constraints   given by $D_a\gamma^a{_{IJ}}= \gamma{_{IJ}}$. Therefore, there are 24 independent first class constraints, this allow us to conclude that the Second Chern invariant is devoid of degrees of freedom and defines a topological field theory too. \\
It is important to note that while in Chern-Simons theory there are present second class constraints  in Pontryagin  there are not. Thus, Pontryagin theory preserves the topological symmetry with only first class constraints and the reducibility condition (\ref{eq36}). \\
With all these results at hand,  we can identify the extended action which is given by 
\begin{eqnarray}
S_E[A_\alpha{^{IJ}}, \Pi^{\alpha}{_{IJ}}, \lambda_0{^{IJ}}, \lambda_a{^{IJ}}, \lambda^{IJ}] & = & \int dx^{4} \Big[ \Pi^{0}{_{IJ}} \dot{A}_0{^{IJ}} + \Pi^{0}{_{IJ}} \dot{A}_0{^{IJ}} - H \nonumber \\
 &  -& \ \lambda_0{^{IJ}} \gamma^{0}{_{IJ}} - \lambda_a{^{IJ}} \gamma^{a}{_{IJ}} - \lambda^{IJ} \gamma_{IJ} \Big], 
\label{eq38}
\end{eqnarray}
where $H=-A_0{^{IJ}} D_a \Pi^{a}{_{IJ}} = -A_0{^{IJ}}\gamma_{IJ}$, and  is a linear combination of Gauss  constraint. 
From the extended action we can identify the extended Hamiltonian  given by 
\begin{equation}
H_E = H + \lambda_0{^{IJ}} \gamma^{0}{_{IJ}} + \lambda_a{^{IJ}} \gamma^{a}{_{IJ}} + \lambda^{IJ} \gamma_{IJ}. 
\label{eq39}
\end{equation}
where is a linear combination of first class constraints. Now,  we shall compute the equations of motion obtained from the extended action (\ref{eq38}), which are given by 
\begin{eqnarray}
\delta A_0{^{IJ}} : \dot{\Pi}^{0}{_{IJ}} & = & \gamma_{IJ}, \nonumber \\
\delta \Pi^{0}{_{IJ}} : \dot{A}_0{^{IJ}} & = & \lambda_0{^{IJ}}, \nonumber \\
\delta A_a{^{IJ}} : \dot{\Pi}^{a}{_{IJ}} & = & (A_{0I}{^{K}} - \lambda_{I}{^{K}}) \Pi^{a}{_{JK}} - (A_{0J}{^{K}} - \lambda_{J}{^{K}}) \Pi^{a}{_{IK}} + 2\alpha \eta^{abc} D_b \lambda_{cIJ}, \nonumber \\
\delta \Pi^{a}{_{IJ}} : \dot{A}_a{^{IJ}} & = & D_a \left(A_0{^{IJ}} - \lambda^{IJ} \right) + \lambda_a{^{IJ}}, \nonumber \\
\delta \lambda_0{^{IJ}} : \gamma^{0}{_{IJ}} & = & 0, \nonumber \\
\delta \lambda_a{^{IJ}} : \gamma^{a}{_{IJ}} & = & 0,  \nonumber \\
\delta \lambda{^{IJ}} : \gamma_{IJ} & = & 0.
\label{eq40}
\end{eqnarray}
\newline
\noindent \textbf{II.I Gauge generator }\\[1ex]
As we have showed, our theory presents a set of first class constraints. In this manner, we will have the presence of gauge transformations.   We  will proceed  to identify the gauge transformations  for the system  by applying the  Castellani's algorithm,  constructing   the follow gauge generator  
\begin{equation}
G = \int dx^{3} \left[ D_0\varepsilon_0{^{IJ}} \gamma^{0}{_{IJ}} + \varepsilon_a{^{IJ}} \gamma^{a}{_{IJ}} + \varepsilon^{IJ} \gamma_{IJ} \right].
\label{eq41}
\end{equation}
Thus, the gauge transformations on  the phase space  are given by 
\begin{eqnarray}
\delta A_0{^{IJ}} & = & D_0 \varepsilon_0{^{IJ}}, \nonumber \\
\delta A_a{^{IJ}} & = & \varepsilon_a{^{IJ}} - D_a \varepsilon^{IJ}, \nonumber \\
\delta \Pi^{0}{_{IJ}} & = & -\varepsilon_I{^{L}} \Pi^{0}{_{LJ}} + \varepsilon_J{^{L}} \Pi^{0}{_{LI}}, \nonumber \\
\delta \Pi^{a}{_{IJ}} & = & \alpha \eta^{abc} D_b \varepsilon_{cIJ} + \Pi^{a}{_{IK}} \varepsilon_J{^{K}} - \Pi^{a}_{JK} \varepsilon_I{^{K}}.
\label{eq41a}
\end{eqnarray}
We can see that in correspondence with the Chern-Simons theory, diffeomorphisms are not present in these gauge transformations. However, we introduce   a set of new gauge parameters  $\varepsilon_0{^{IJ}}=\varepsilon{^{IJ}}= -\xi^\rho A_\rho{^{IJ}}$ and   $\varepsilon_\mu{^{IJ}}=-\xi^\rho F_{\rho \mu}^{IJ}$,  allowing us  rewrite  the gauge transformations as 
\begin{equation}
A'_\mu {^{IJ}} \rightarrow A_\mu {^{IJ}}+ {\mathcal{L}}_\xi  A_{\mu}{^{IJ}}, 
\label{eq42}
\end{equation} 
which corresponds to diffeomorphisms. Therefore diffeomorphisms corresponds to an internal symmetry of the theory. It is important to observe,  that  the Pontryagin invariant which is defined in four dimensions inherit the principal symmetries of Chern-Simons theory defined in three  dimensions, as for instance the   invariance  under diffeomorphisms. In this manner, because of  Pontryagin is defined in four dimensions  its   quantization study  could be a good  attempt to understand the constrained gravitational field  because we have at hand similar  symmetries. However,  we need to be careful because    Pontryagin invariant is a  topological field  theory such as has been   showed in this section   while general relativity is not,  because there exists two degrees of freedom per point of the space \cite{18}. 
\newline
\newline
\newline
\noindent \textbf{III. Hamiltonian dynamics for  modified Pontryagin invariant }\\[1ex]
We will  complete the analysis of this work by  performing  a pure Dirac  analysis for a  modified version of  Pontryagin theory.  In particular, we shall  reproduce the results discussed  above. \\ 
For our purposes,  we will work  with  the next action \cite{11}
\begin{equation}
S[A, R]= \alpha \int_M \frac{1}{2}R^{IJ}\wedge R_{IJ}- R_{IJ}\wedge(dA^{IJ} + A{^{I}}_K\wedge A{^{KJ}} ). 
\label{eq43}
\end{equation}
Now, we will consider that the  1-form $A^{IJ}$ and the two-form $R^{IJ}$ represents our new independent set of dynamical  variables. We can see  with  this  election of variables,  that we have   extended the configuration  space respect to Pontryagin theory   and therefore, by performing  the Hamiltonian analysis   we will extend the phase space. The equations of motion obtained from the action  (\ref{eq43}) are given by 
\begin{eqnarray}
R^{IJ}&= &dA^{IJ} + A{^{I}}_K\wedge A{^{KJ}},  \nonumber \\
DR^{IJ}&=&0. 
\label{eq44}
\end{eqnarray}
 By using  the  equation of motion  (\ref{eq44}) in  (\ref{eq43}) we can eliminate $R$, obtaining the same action (\ref{eq26}) and the equations of motion (\ref{eq26a}) \cite{18}. In this manner,  the following  question rise; will be  the same symmetries  for the action (\ref{eq43})  those  found  above for the action (\ref{eq26})?. Our answer  at Lagrangian level   can be  yes. However, at Hamiltonian level  we need to be careful because of   two systems sharing the same  equations of motion,  not necessary  yields to the same symmetries and symplectic structures \cite{11} (see  \cite{19} as well).  Therefore, we will answer the question by  performing  a pure Dirac method for the action (\ref{eq43} ) and then,  compare the results with those obtained above for Pontryagin theory. \\
From now  on,  to avoid confusion with   Pontryagin action,   we will refer  to the action (\ref{eq43}) as modified Pontryagin theory.   \\
By performing the 3+1 decomposition  for the modified  action (\ref{eq43}) we find 
\begin{equation}
S[A,R]= \int_\Sigma dx^3\int dt\left[\frac{\alpha}{2} \eta^{abc}R_{0aIJ} \left( F^{IJ}{_{bc}}- R{_{bc}}^{IJ}\right)  + \frac{\alpha}{2}  \eta^{abc}R{_{bc}}^{IJ}\left( \dot{A}_a{^{IJ}} -D_a A_0{^{IJ}}\right) \right], 
\label{eq45}
\end{equation}
where we identify with $F_{abIJ }= \partial_a A_{bIJ} -\partial_b A_{aIJ} + A_{aI}{^{L}}A_{bLJ}-A_{bI}{^{L}}A_{aLJ} $  the two-form curvature.   \\
For this modified theory we have a set of  $(A_{\alpha}{^{IJ}}, R{^{IJ}}_{\alpha \beta})=60$ dynamical variables, so  Dirac's method calls for  the definition of the momenta $(\Pi^{\alpha}{_{IJ}}, \Pi^{\mu \nu}{_{IJ}})$ canonically conjugate to $(A_{\alpha} {^{IJ}}, R^{IJ}{_{\mu \nu}})$
\begin{eqnarray}
\Pi^{\alpha}{_{IJ}}&=& \frac{\delta {\mathcal{L}} }{ \delta \dot{A}_{\alpha} {^{IJ}} } ,\nonumber \\
\Pi^{\mu \nu}{_{IJ}}&=& \frac{\delta {\mathcal{L}} }{ \delta \dot{R}{_{\mu \nu}}^{IJ} }.
\label{eq46}
\end{eqnarray}
The matrix elements of the Hessian 
\begin{equation}
\frac{\partial^2{\mathcal{L}} }{\partial( \partial_\mu (A_{\alpha} {^{IJ}} )) \partial(\partial_\mu (A_{\beta} {^{IJ}} )) },\quad \quad \frac{\partial^2{\mathcal{L}} }{\partial( \partial_\mu (A_{\alpha} {^{IJ}} ) )\partial(\partial_\mu (R{_{\rho \nu}}^{IJ} ) ) },  \quad \quad \frac{\partial^2{\mathcal{L}} }{\partial( \partial_\mu (R{_{\rho \nu}}^{IJ} )) \partial(\partial_\mu (R{_{\gamma \sigma}}^{IJ}) ) }, 
\label{eq47}
\end{equation}
are identically zero, the rank of the Hessian is zero, thus, we expect 60 primary constraints. From the definition of the momenta (\ref{eq46}) we identify the next 60 primary constraints 
\begin{eqnarray}
\phi^{0}{_{IJ}}&:=& \Pi^{0}{_{IJ}} \approx 0 ,\nonumber \\
\phi^{a}{_{IJ}}&:=& \Pi^{a}{_{IJ}} - \frac{\alpha}{2}  \eta^{abc}R{_{bcIJ}}\approx 0, \nonumber \\
\phi^{0a}{_{IJ}}&:=& \Pi^{0a}{_{IJ}} \approx 0, \nonumber \\ 
\phi^{ab}{_{IJ}}&:=& \Pi^{ab}{_{IJ}} \approx 0.
\label{eq48}
\end{eqnarray}
 For the system under study, the canonical Hamiltonian  is given by 
\begin{equation}
H_{c}= \int  dx^3 \left[ -\frac{1}{2} A_{0}{^{IJ}} D_a\Pi^{a}{_{IJ}}  + R{_{0a}}^{IJ}\left( \Pi^{a}{_{IJ}} - \frac{\alpha}{2} \eta^{abc} F_{ bcIJ}\right) \right].
\label{eq49}
\end{equation}
In this manner, with the canonical Hamiltonian and the primary constraints at hand, can be identify the primary Hamiltonian expressed  by 
\begin{equation}
H_P= H_c + \int dx^3 \left[  \lambda^{IJ}{_{0}} \phi{_{IJ}}^{0} + \lambda^{IJ}{_{a}} \phi{_{IJ}}^{a}+\lambda_{0a}{^{IJ}}\phi^{0a}{_{IJ}} +\lambda_{ab}{^{IJ}}\phi^{ab}{_{IJ}}  \right],
\label{eq50} 
\end{equation}
where $ \lambda^{IJ}{_{0}}$,  $\lambda^ {IJ}{_{a}}, \lambda_{0a}{^{IJ}}$ and $ \lambda_{ab}{^{IJ}}$  are Lagrange multipliers enforcing the constraints.\\
 For the theory under study, can be identified the next non-vanishing fundamental brackets 
\begin{eqnarray}
\{ A_{\alpha} {^{IJ}}(x),\Pi^{\beta}{_{KL}}(y) \}& =& \frac{1}{2} \delta {^ \beta} _\alpha \left( \delta^I {_K} \delta^J{_L} - \delta^I{ _L} \delta^J{_K } \right) \delta^3(x-y), \nonumber \\
\{ R {_{\mu \nu}}^{IJ}(x), \Pi^{\alpha \beta}{_{KL}}(y) \}& =& \frac{1}{4}\left( \delta {^ \alpha} _\mu \delta^\beta{_{\nu}}- \delta {^ \alpha} _\nu \delta^\beta{_{\mu}} \right)\left( \delta^I {_K} \delta^J{_L} - \delta^I{ _L} \delta^J{_K } \right) \delta^3(x-y).
\label{eq51}
\end{eqnarray}
Now,  we need to identify if our modified theory presents secondary constraints. For this aim, we compute the  60 $\times$ 60 matrix whose entries are the Poisson brackets among the primary constraints (\ref{eq48})  
\begin{eqnarray}
\{\phi{_{IJ}}^{0} (x),\phi{_{KL}}^{0} (y) \} &=& 0, \nonumber \\
 \{\phi{_{IJ}}^{0} (x),\phi^a{_{KL}} (y) \} &=& 0, \nonumber \\
  \{\phi{_{IJ}}^{0} (x),\phi^{0a}{_{KL}} (y) \}&=& 0, \nonumber \\
    \{\phi{_{IJ}}^{0} (x),\phi^{ab}{_{KL}} (y) \}&=& 0, \nonumber \\
\{\phi{_{IJ}}^{a} (x),\phi^{a}{_{KL}}(y) \} &=& 0, \nonumber \\ 
\{\phi{_{IJ}}^{a} (x),\phi^{0a}{_{KL}} (y) \} &=& 0, \nonumber\\
\{\phi{_{IJ}}^{a} (x),\phi^{cd}{_{KL}} (y) \} &=& -\frac{\alpha}{4} \eta^{acd}\left(\eta_{IK}\eta_{JL}-\eta_{IH}\eta_{JF}     \right) \delta^3(x-y), \nonumber\\
\{\phi{_{IJ}}^{0a} (x),\phi^{0b}{_{KL}} (y) \} &=& 0, \nonumber\\
\{\phi{_{IJ}}^{0a} (x),\phi^{cd}{_{KL}} (y) \} &=& 0, \nonumber\\
\{\phi{_{IJ}}^{ab} (x),\phi^{cd}{_{KL}} (y) \} &=& 0, \nonumber\\
\label{eq52}
\end{eqnarray}
this matrix has rank= 36 and 24  linearly independent null-vectors.  Consistency  and the null vectors yields to identify  the next 24 secondary constraints 
\begin{eqnarray}
\dot{\phi}^{0}{_{IJ}}&=& \{\phi^{0}{_{IJ}} (x), {H}_{P} \} \approx 0 \quad \Rightarrow \quad \psi_{IJ}:= D_a \Pi^a{_{IJ}} \approx 0. \nonumber \\
\dot{\phi}^{0a}{_{IJ}}&= &\{\phi^{0a}{_{IJ}} (x), {H}_{P} \} \approx 0 \quad \Rightarrow \quad \psi{^{0a}}_{IJ}:= \Pi{^{a}}_{IJ}-\frac{\alpha}{2}\epsilon^{abc}F_{bcIJ}  \approx 0, 
\label{eq53}
\end{eqnarray}
and the next values for the Lagrange multipliers 
\begin{eqnarray}
\dot{\phi}^a{_{IJ}}= \{\phi^a{_{IJ}} (x), {H}_{P} \} \approx 0 \quad & \Rightarrow& \quad \frac{1}{2}\left[\Pi^a{_{JL}}\eta_{KI}-\Pi^a{_{IL}}\eta_{KJ}  \right] A_0{^{KL}}- \alpha \eta^{abi} D_i R_{0bIJ}  \nonumber \\ &-&\frac{\alpha}{2} \eta^{acd} \lambda_{cdIJ}\approx 0, \nonumber \\
\dot{\phi}^{ab}{_{IJ}}= \{\phi^{ab}{_{IJ}} (x), {H}_{P} \} \approx 0 \quad &\Rightarrow& \quad  \eta^{abc} \lambda_{cIJ}  \approx 0. 
\label{eq54}
\end{eqnarray}
Consistency requires that the conservation in the time of the constraints vanish as well. For this theory there no, third constraints.  At this point, we need to identify from   primary and secondary constrains which ones corresponds to  first and second class. For this purpose,  we need to calculate the rank and the null-vectors of the  84$\times$ 84 matrix whose entries will be the Poisson brackets among  primary and secondary constraints, this is 
\begin{eqnarray}
\{\phi^0{_{IJ}} (x),\phi^0{_{KL}} (y) \} &=& 0, \nonumber \\
 \{\phi^0{_{IJ}} (x),\phi^a{_{KL}} (y) \} &=& 0, \nonumber \\
  \{\phi^0{_{IJ}} (x),\phi^{0a}{_{KL}} (y) \}&=& 0, \nonumber \\
    \{\phi^0{_{IJ}} (x),\phi^{ab}{_{KL}} (y) \}&=& 0, \nonumber \\
        \{\phi^0{_{IJ}} (x),\psi{_{KL}} (y) \}&=& 0, \nonumber \\
            \{\phi^0{_{IJ}} (x),\psi^{0a}{_{KL}} (y) \}&=& 0, \nonumber \\
\{\phi^a{_{IJ}} (x),\phi^{a}{_{KL}}(y) \} &=& 0, \nonumber \\ 
\{\phi^a{_{IJ}} (x),\phi^{0a}{_{KL}} (y) \} &=& 0, \nonumber\\
\{\phi^a{_{IJ}} (x),\phi^{cd}{_{KL}} (y) \} &=& -\frac{\alpha}{4} \eta^{acd}\left(\eta_{IK}\eta_{JL}-\eta_{IL}\eta_{JK}     \right) \delta^3(x-y), \nonumber\\
\{\phi^a{_{IJ}} (x),\psi{_{KL}} (y) \}&=& -\frac{1}{2}\left[ \Pi^{a}{_{JL}} \eta_{KI} -\Pi^{a}{_{IL}} \eta_{KJ} +\Pi^{a}{_{KJ}} \eta_{LI}-\Pi^{a}{_{KI}} \eta_{LJ}    \right] \delta^3(x-y), \nonumber \\
\{\phi^a{_{IJ}} (x),\psi^{0b}{_{KL}} (y) \}&=& \frac{\alpha}{2} \eta^{abc}  \Big\{ \partial_c \delta^3(x-y) \left(\eta_{KI}\eta_{LJ}-\eta_{KJ}\eta_{LI} \right)+  \left(\omega_{cIL}\eta_{KJ}- \omega_{cJL}\eta_{KI} \right)\delta^3(x-y)  \nonumber \\ &+& \left(\omega_{cKI} \eta_{LJ} - \omega_{cKJ} \eta_{LI}\right)\delta^3(x-y) \Big \} , \nonumber \\
\{\phi^{0a}{_{IJ}} (x),\phi^{0b}{_{KL}} (y) \} &=& 0, \nonumber\\
\{\phi^{0a}{_{IJ}} (x),\phi^{cd}{_{KL}} (y) \} &=& 0, \nonumber\\
\{\phi^{0a}{_{IJ}} (x),\psi{_{KL}} (y) \}&=& 0, \nonumber \\
\{\phi^{0a}{_{IJ}} (x),\psi^{0b}{_{KL}} (y) \}&=& 0, \nonumber \\
\{\phi^{ab} {_{IJ}}(x),\phi^{cd}{_{KL}} (y) \} &=& 0, \nonumber\\
\{\phi^{ab} {_{IJ}} (x),\psi{_{KL}} (y) \}&=& 0, \nonumber \\
\{\phi^{ab} {_{IJ}} (x),\psi^{0c}{_{KL}} (y) \}&=& 0, \nonumber \\
\{\psi{_{IJ}} (x),\psi{_{KL}} (y) \}&=&  \frac{1}{2} \left(\psi{_{IJ}}\eta_{KI} +\psi{_{JK}}\eta_{LI} +\psi{_{IL}}\eta_{KJ} +\psi{_{KI}}\eta_{LJ}   \right) \delta^3(x-y)\approx0, \nonumber \\
\{\psi{_{IJ}} (x),\psi^{0a}{_{KL}} (y) \}&=&  \frac{1}{2} \left(\psi^{0a}{_{LJ}}\eta_{KI} +\psi^{0a}{_{JK}}\eta_{LI} +\psi^{0a}{_{IL}}\eta_{KJ} +\psi^{0a}{_{KI}}\eta_{LJ}   \right) \delta^3(x-y)\approx0, \nonumber \\
\{\psi^{0a}{_{IJ}} (x),\psi^{0b}{_{KL}} (y) \}&=& 0,
\label{eq55}
\end{eqnarray}
this matrix has rank=36 and 48 null-vectors. From the null vectors we can   identify the next  48  first class constraints 
\begin{eqnarray}
\gamma^0{_{IJ}} &:=& \Pi^{0}{_{IJ}}  \approx 0 , \nonumber \\
\gamma^{0a}{_{IJ}} &:=& \Pi^{0a}{_{IJ}}  \approx 0 , \nonumber \\
\gamma{_{IJ}} &:=& D_a\Pi^a{_{IJ}} - \left(\Pi^{ab}{_{I}}^F R_{abFJ }-\Pi^{ab}{_{J}}^F R_{ abFI} \right)  \approx 0,\nonumber \\
\gamma^{0a}{_{IJ}} &:=& \Pi{^{a}}_{IJ}-\frac{\alpha}{2} \eta^{abc}F_{bcIJ}  + 2 D_b\Pi^{ab}{_{IJ}}   \approx 0,
\label{eq56}
\end{eqnarray}
We can observe, that the third equation of (\ref{eq56}) can be identified as the Gauss constraint for this extended Pontryagin theory.  On the other hand, the rank of the matrix  (\ref{eq55}) yields to  identify the following t 36 second class constraints 
\begin{eqnarray}
\chi^a{_{IJ}} &:=& \Pi^{a}{_{IJ}} - \frac{\alpha}{2}  \eta^{abc}R{_{bcIJ}}\approx 0, \nonumber \\
\chi^{ab}{_{IJ}} &:=& \Pi^{ab} {_{IJ}} \approx 0. 
\label{eq57}
\end{eqnarray}
The  identification of  first and  second class constraints will allow us to carry out the counting of degrees of freedom; we have 120  canonical variables  $(A_{\alpha} {^{IJ}},R{_{\mu \nu}}^{IJ}, \Pi^{\alpha}{_{IJ}},\Pi^{\mu \nu}{_{IJ}} )$, 48  first class constraints $(\gamma^0{_{IJ}},\gamma^{0a}{_{IJ}} , \gamma{_{IJ}},\gamma^{0a}{_{IJ}} )$  and 36 second class constraints $(\chi^a{_{IJ}},\chi^{ab}{_{IJ}} )$. However, just as for  Pontryagin theory   Bianchi's identities $DF=0$ implies  6 reducibility conditions among the first class constraints.  We can see that  for the   modified Pontryagin theory,     reducibility conditions   has  a   longer  expression  than  Pontryagin  (see (\ref{eq36})):  $D_a\gamma^{0a}{_{IJ}} - \gamma{_{IJ}}- \left (\chi^{ab}{_{I}}^F R_{abFJ }-\chi^{ab}{_{J}}^F R_{abFI } \right)  -2D_aD_b\chi^{ab}{_{IJ}}= 0$. Therefore,   we have 42 independent first class constraints. By using this fact,   the counting of degrees of  freedom   yields to conclude that this modified Pontryagin  theory is devoid of degrees of  freedom and   defines a topological field theory too.  It is important to remark that we can reproduce the results found for the action (\ref{eq26}) by considering the second class  constraints (\ref{eq57}) as strong identities, thus, the constraints (\ref{eq56}) will be reduced  to (\ref{eq37}).     \\
By following with the method, we need to compute the  algebra of  constraints, for this fact  it is convenient  rewrite them  as 
\begin{eqnarray}
\phi_1 &:=& \gamma^0{_{IJ}} \left[ A \right]= \int dx^3  A^{IJ} \left[ \Pi^0{_{IJ}} \right], \nonumber \\
\phi_2  &:=&\gamma^{0a}{_{IJ}} \left[ B \right]= \int dx^3 B_{0a}{^{IJ}} \left[  \Pi^{0a}{_{IJ}}  \right], \nonumber \\
\phi_3  &:=& \gamma{_{IJ}} \left[ C \right]= \int dx^3 C_a{^{IJ}} \left[D_a  \Pi^a{_{IJ}} - \left(\Pi^{ab}{_{I}}^F R_{FJ ab}-\Pi^{ab}{_{J}}^F R_{FI ab} \right) \right], \nonumber \\
\phi_4 &:=& \gamma^{0a}{_{IJ}} \left[ \mathbf{D} \right]= \int dx^3  {\mathbf{D}}{_{0a}}^{IJ} \left[  \Pi{^{a}}_{IJ}-\frac{\alpha}{2} \epsilon^{abc}F_{bcIJ}  + 2 D_b\Pi^{ab}{_{IJ}} \right], \nonumber \\
\phi_5  &:=&\chi^a{_{IJ}}  \left[ F \right]= \int dx^3 F_a{^{IJ}} \left[ \Pi^{a}{_{IJ}} - \frac{\alpha}{2}  \eta^{abc}R^{IJ}{_{bc}}  \right], \nonumber \\
\phi_6  &:=&\chi^{ab}{_{IJ}} \left[ G \right]= \int dx^3 G{_{ab}}{^{IJ}} \left[ \Pi^{ab} {_{IJ}}  \right].
\label{eq58}
\end{eqnarray} 
Thus, the algebra of   constraints is given  by 
\begin{eqnarray}
\Big\{ \phi_1\left[ A^{IJ} \right], \phi_1\left[ A'^{KL} \right] \Big\} &=& 0, \nonumber \\
\Big\{\phi_1\left[ A^{IJ} \right],\phi_2 \left[ B_{0a}{^{KL}} \right] \Big\} &=& 0, \nonumber \\
\Big\{\phi_1\left[ A^{IJ} \right], \phi_3\left[ C{^{KL}}  \right] \Big\}&=& 0, \nonumber \\
\Big\{\phi_1\left[ A^{IJ} \right], \phi_4 \left[ {\mathbf{D}}{_{0a}}^{KL} \right] \Big\}&=& 0, \nonumber \\
\Big\{\phi_1\left[ A^{IJ} \right],\phi_5   \left[ F_a{^{KL}} \right] \Big \}&=& 0, \nonumber \\
\Big\{\phi_1\left[ A^{IJ} \right], \phi_6 \left[ G{_{ab}}{^{KL}} \right]  \Big\}&=& 0, \nonumber \\
\Big\{\phi_2 \left[ B_{0a}{^{IJ}}  \right] ,\phi_2 \left[ B'_{0b}{^{KL}} \right] \Big\} &=& 0, \nonumber \\
\Big\{\phi_2 \left[ B_{0a}{^{IJ}}  \right] , \phi_3\left[ C{^{KL}}  \right] \Big\}&=& 0, \nonumber \\
\Big\{\phi_2 \left[ B_{0a}{^{IJ}}  \right] , \phi_4 \left[ {\mathbf{D}}{_{0b}}^{KL} \right] \Big\}&=& 0, \nonumber \\
\Big\{\phi_2 \left[ B_{0a}{^{IJ}}  \right] ,\phi_5   \left[ F_b{^{KL}} \right] \Big \}&=& 0, \nonumber \\
\Big\{\phi_2 \left[ B_{0a}{^{IJ}}  \right] , \phi_6 \left[ G{_{cd}}{^{KL}} \right]  \Big\}&=& 0, \nonumber \\
\Big\{\phi_3\left[ C{^{IJ}} \right],\phi_3\left[ C'{^{KL}} \right]  \Big\} &=& \int dx^3\left[C^{IK}C'_K{^{J}}-C^{JK}C'_K{^{I}} \right]  \gamma{_{IJ}} \approx0, \nonumber \\
\Big\{\phi_3\left[ C{^{IJ}} \right],\phi_4\left[ {\mathbf{D}}{_{0a}}^{KL} \right]   \Big\} &=& \int dx^3\left[C^{IK}{\mathbf{D}}_{0aK}{^{J}}-C^{JK}{\mathbf{D}}_{0aK}{^{I}} \right]  \gamma^{0a}{_{IJ}} \approx0, \nonumber \\
\Big\{\phi_3\left[ C{^{IJ}} \right],\phi_5\left[ F_a{^{KL}} \right]  \Big\} &=& \int dx^3\left[C^{IK}F_{aK}{^{J}}-C^{JK}F{_{aK}}{^{I}} \right] \chi^a{_{IJ}}  \approx 0, \nonumber \\
\Big\{\phi_3\left[ C{^{IJ}} \right],\phi_6\left[ G{_{ab}}{^{KL}}  \right] \Big\} &=&  0, \nonumber \\
\Big\{\phi_4 \left[ {\mathbf{D}}_{0a}{^{IJ}}  \right] , \phi_4 \left[ {\mathbf{D}}'_{0b}{^{KL}}  \right]  \Big\}&=& 0, \nonumber \\
\Big\{\phi_4 \left[ {\mathbf{D}}_{0a}{^{IJ}}  \right] , \phi_5 \left[ F_{b}{^{KL}}  \right]  \Big\}&=& 0, \nonumber \\
\Big\{\phi_4 \left[ {\mathbf{D}}_{0a}{^{IJ}}  \right] , \phi_6 \left[ G'_{cd}{^{KL}}  \right]  \Big\}&=& 0, \nonumber \\
\Big\{\phi_5 \left[ F_{a}{^{IJ}}  \right] , \phi_5 \left[ F'_{a}{^{KL}}  \right]  \Big\}&=& 0, \nonumber \\
\Big\{\phi_5 \left[ F_{a}{^{IJ}}  \right] , \phi_6 \left[ G'_{ab}{^{KL}}  \right]  \Big\}&=& -\frac{\alpha}{4}\eta^{aij}\int dx^3 \left[ F_{aKH}G{_{ij}}^{KH}-F_{iKH}G{_{aj}}^{KH} \right], \nonumber \\
\Big\{\phi_6 \left[ G_{ab}{^{IJ}}  \right] , \phi_6 \left[ G'_{cd}{^{KL}}  \right]  \Big\}&=& 0, \nonumber \\
\label{eq59}
\end{eqnarray}
where we can see clearly that  (\ref{eq56}) and (\ref{eq57}) form a first and second class constraints set respectively. It is important to observe,   that the algebra among  the constraints for this modified Pontryagin theory shares a closed relation with the constraint algebra for the Chern-Simons theory (\ref{eq18}) (see   the Poisson's brackets between  $\phi_3,\phi_4, \phi_5$ of (\ref{eq59}) and $\phi_2,\phi_3$ of (\ref{eq18})). In addition,  now this modified theory presents second class constraints as well.\\
With all these results at hand, we can use the Lagrange's multipliers values (\ref{eq54}), the first class constraints (\ref{eq56}) and  the second class constraints (\ref{eq57}) to  identify the extended  action for the theory expressed by 
\begin{eqnarray}
&S_E&\left[ A_{\alpha} {^{IJ}},\Pi^{\alpha}{_{IJ}}, R{_{\mu \nu}}^{IJ}, \Pi^{\mu \nu}{_{IJ}}, u_0{^{IJ}}, u_{0a}{^{IJ}}, u{^{IJ}}, u{_{a}}^{IJ}, v_a{^{IJ}}, v{_{ab}}{^{IJ}}  \right] =\int \Big\{ \dot{A}_{\alpha} {^{IJ}}\Pi^{\alpha}{_{IJ}}+ \dot{ R}{_{0a}}^{IJ}\Pi^{0a}{_{IJ}}\nonumber \\ 
&+& \dot{ R}{_{ab}}^{IJ}\Pi^{ab}{_{IJ}} -  H-u_0{^{IJ}} \gamma^0{_{IJ}}- u_{0a}{^{IJ}} \gamma^{0a}{_{IJ}} -    u{^{IJ}} \gamma{_{IJ}}-     u{_{a}}^{IJ}\gamma^{0a}{_{IJ}}    -v_a{^{IJ}} \chi^a{_{IJ}}     -v{_{ab}}{^{IJ}} \chi^{ab}{_{IJ}}         \Big\} dx^4, \nonumber \\
\end{eqnarray}
where  $H$ is  only linear  combination of  first class constraints 
\begin{equation}
H= \frac{1}{2}A_0{^{IJ}}\left[D_a \Pi^a{_{IJ}}  - \left(\Pi^{ab}{_{I}}^F R_{abFJ }-\Pi^{ab}{_{J}}^F R_{ abFI} \right)  \right]- R_{0a}{^{IJ}}\left[ \Pi{^{a}}_{IJ}-\frac{\alpha}{2} \epsilon^{abc}F_{bcIJ}  + 2 D_b\Pi^{ab}{_{IJ}} \right],
\label{eq60}
\end{equation}
and $u_0{^{IJ}}, u_{0a}{^{IJ}}, u{^{IJ}}, u{_{a}}^{IJ}, v_a{^{IJ}}, v{_{ab}}{^{IJ}}$ are the Lagrange multipliers enforcing the first and second class  constraints. We can observe, that   by considering the second class constraints as strong equations   the Hamiltonian (\ref{eq60}) is reduced to that Hamiltonian quantized in \cite{11} where was performed the Hamiltonian analysis on a smaller phase space. In this manner,  we have here a best description at classical level than that reported in \cite{11}.   \\
From the extended action we can identify the extended Hamiltonian  which is given by
\begin{equation}
H_E= H-u_0{^{IJ}} \gamma^0{_{IJ}}- u_{0a}{^{IJ}} \gamma^{0a}{_{IJ}} - u{^{IJ}} \gamma{_{IJ}}- u{_{a}}^{IJ}\gamma^{0a}{_{IJ}}.
\end{equation}
As we Know, the equations of motion obtained by means of the extended Hamiltonian in general are mathematically different   with the Euler-Lagrange equations, but the difference is unphysical \cite{6}.\\
We will continue  this section computing   the equations of motion obtained  from the extended action. The equations of motion derived from the extended action are given by 
\begin{eqnarray}
\delta A{_{0}}^{IJ}:  \dot{\Pi}^0{_{IJ}}&=&\frac{1}{2}\left[ D_a\Pi^a{_{IJ}} - \left(\Pi^{ab}{_{I}}^F R_{abFJ }-\Pi^{ab}{_{J}}^F R_{ abFI} \right)\right] ,  \nonumber \\
\delta \Pi^0{_{IJ}}:  \dot{A}{_{0}}^{IJ} &=& u{_{0}}^{IJ},  \nonumber \\
\delta A{_{a}}^{IJ}: \dot{\Pi}^a{_{IJ}}&=&\left[A_{0J}{^{F}} + u_J{^{F}} \right] \Pi^a{_{IF}}-\left[A_{0I}{^{F}} + u_I{^{F}} \right] \Pi^a{_{JF}} - \alpha \eta^{abc}\left[D_bR_{0cIJ}-D_bu_{cIJ} \right]  \nonumber \\ &+& 2\left[u_{bI}{^{F}}- R_{0bI}{^{F}}  \right]\Pi^{ab}{_{JF}}-2\left[u_{bJ}{^{F}}- R_{0bJ}{^{F}}  \right]\Pi^{ab}{_{IF}} , \nonumber \\
\delta \Pi^a{_{IJ}}:   \dot{A}{_{a}}^{IJ}&=&-D_a\left(\frac{1}{2}A_0{^{IJ}}+u^{IJ}  \right) + \left(u_a{^{IJ}} -R_{0a}{^{IJ}} \right)+ v_a{^{IJ}}, \nonumber \\
\delta R{_{0a}}^{IJ}:  \dot{\Pi}^{0a}{_{I}}&=& - \left[ \Pi{^{a}}_{IJ}- \frac{\alpha}{2} \eta^{abc}F_{bcIJ}  + 2 D_b\Pi^{ab}{_{IJ}} \right] ,  \nonumber \\
\delta \Pi^{0a}{_{IJ}}:  \dot{R}{_{0a}}^{IJ} &=&  u{_{0a}}^{IJ},  \nonumber \\
\delta R{_{ab}}^{IJ} : \dot{\Pi}^{ab}{_{IJ}}&=& \left[ \frac{1}{2}A_0{^{F}}_{J} + u^{F}{_{J}} \right] \Pi^{ab}{_{FI}} -  \left[ \frac{1}{2}A_0{^{F}}_{I} + u^{F}{_{I}} \right] \Pi^{ab}{_{FJ}} + \frac{\alpha}{2}\eta ^{abc}v_{cIJ}, \nonumber \\
\delta \Pi^{ab}{_{IJ}}:  \dot{R}{_{ab}}^{IJ}&=&  \left[ \frac{1}{2}A_0 {^{JF}}+ u^{JF}  \right]R_{ab}{^{I}}_F- \left[ \frac{1}{2}A_0 {^{IF}}+ u^{IF}  \right]R_{ab}{^{J}}_F + D_a\left( u_b{^{IJ}}-R_{0b}{^{IJ}} \right) \nonumber \\  &-& D_b\left( u_a{^{IJ}}-R_{0a}{^{IJ}} \right)+ v_{ab}{^{IJ}}, \nonumber \\
\delta u_0{^{IJ}} : \gamma^0{_{IJ}}&=&0, \nonumber \\
\delta  u_{0a}{^{IJ}}: \gamma^{0a}{_{IJ}}&=&0, \nonumber \\
\delta  u{^{IJ}}: \gamma{_{IJ}}&=&0, \nonumber \\
\delta u{_{a}}^{IJ}:\gamma^{0a}{_{IJ}} &=&0, \nonumber \\
\delta v_a{^{IJ}} :\chi^a{_{IJ}} &=&0, \nonumber \\
\delta v{_{ab}}{^{IJ}}: \chi^{ab}{_{IJ}}                                              &=&0 . 
\label{eq61}
\end{eqnarray}
\newline
\noindent \textbf{III.I Gauge generator }\\[1ex]
As we have showed, our modified theory presents a set of first class constraints.  Therefore,  we need to identify the form of gauge transformations  generated for these constraints.  For this part,   we will find the gauge transformations generated by the first class constraints (\ref{eq56})  by  using the Castellani's algorithm in essence constructing    the follow gauge generator   
\begin{eqnarray}
G= \int_\Sigma \left[D_0 \varepsilon_0{^{IJ}} \gamma^0{_{IJ}}  + D_0 \varepsilon_{0a}{^{IJ}} \gamma^{0a}{_{IJ}} +\varepsilon^{IJ}\gamma{_{IJ}} + \varepsilon_a {^{IJ}}\gamma^{0a}{_{IJ}}  \right], 
\label{eq62}
\end{eqnarray}
thus, we find the following gauge transformations on the phase  space 
\begin{eqnarray}
\delta_0 A{_{0}}^{IJ} &=&D_0  \varepsilon{_{0}}^{IJ},  \nonumber \\
\delta_0 A{_{a}}^{IJ} &=& -  D_a  \varepsilon^{IJ} + \varepsilon_a{^{IJ}},  \nonumber \\
\delta_0 R{_{0a}}^{IJ} &=&   D_0  \varepsilon{_{0a}}^{IJ},  \nonumber \\
\delta_0 R{_{ab}}^{IJ} &=& \left[ D_a \varepsilon{_{b}}^{IJ} - D_b\varepsilon{_{a}}^{IJ} \right] + \left[ \varepsilon^{IF} R_{abF}{^{J}}- \varepsilon^{JF}R_{abF}{^{I}} \right],  \nonumber \\
\delta_0 \Pi^0{_{IJ}}   &=&   \varepsilon{_{0J}}^{L}\Pi^0{_{IL}}- \varepsilon{_{0I}}^{L}\Pi^0{_{JL}} + \varepsilon{_{0J}}^{L} \Pi^{0a}{_{IL}}- \varepsilon{_{0J}}^{L} \Pi^{0a}{_{JL}},  \nonumber \\
\delta_0 \Pi^a{_{IJ}}   &=& \left[\Pi^a{_{IL}}\varepsilon_J{^{L}}- \Pi^a{_{JL}}\varepsilon_I{^{L}}  \right]+ \alpha\eta^{adc}D_d\varepsilon_{cIJ} + 2\left[ \Pi^{ab}{_{KI}}\varepsilon_b {^{L}}_J- \Pi^{ab}{_{KJ}}\varepsilon_b {^{L}}_I \right],  \nonumber \\
\delta_0 \Pi^{0a}{_{I}}  &=&   0,  \nonumber \\
\delta_0 \Pi^{ab}{_{IJ}}  &=& - \left[\Pi^{ab}{_{IF}} \varepsilon^{F}{_{J}}- \Pi^{ab}{_{JF}} \varepsilon^{F}{_{I}} \right].  
\label{eq63}
\end{eqnarray}
It is important to observe,  that we obtain relevant  results  when   are   considered     the   second class constraints as strong equations. By taking the second class constraints as strong equations,    the   gauge transformations obtained above are reduced to  (\ref{eq41a}) which corresponds to  Pontryagin theory.  On the other hand, with all these results at hand, we can find in particular a close  relation among the  gauge transformations of this modified Pontirjagin theory and the gauge transformations  reported  in  the case of a $BF$ theory \cite{11c}. In fact, the modified version for the Pontryagin theory (equation (\ref{eq43})) has a $BF$ form.   However,   in this work we find   a big difference respect to \cite{11c},    since   in  \cite{11c} we can observe  that there exists   only first class constraints, while in this work we have first and second class kind. The reason for  that  difference  is because in \cite{11c} the Hamiltonian analysis was performed on a smaller phase space context  and a complete analysis  was not reported.  \\
Just as in last section,   we can introduce the new set of  parameters $\varepsilon_0{^{IJ}}=\varepsilon{^{IJ}}= -\xi^\rho A_\rho{^{IJ}}$,  $\varepsilon_\mu{^{IJ}}=-\xi^\rho R_{\rho \mu}^{IJ}$ and taking on  to account the equations of motion (\ref{eq45}),  we find that the gauge transformations take the next form   
\begin{eqnarray}
A'_\mu {^{IJ}}& \rightarrow& A_\mu {^{IJ}}+ {\mathcal{L}}_\xi  A_{\mu}{^{IJ}} + \xi^\rho \left(F_{\rho \mu}{^{IJ}}-R_{\rho \mu}{^{IJ}} \right), \nonumber \\
R'_{\mu \nu}{^{IJ}}& \rightarrow& R_{\mu \nu}{^{IJ}} + {\mathcal{L}}_\xi  R_{\mu \nu}{^{IJ}} + \xi^{\rho} \left[ D_\nu R_{\mu \rho}{^{IJ}} +D_\mu R_{\rho \nu}{^{IJ}}+  D_\rho R_{\nu \mu }{^{IJ}} \right],  
\label{eq64}
\end{eqnarray}
where we can observe that corresponds to diffeomorphisms. In this manner, diffeomorphisms corresponds to an internal symmetry for the theory. It is important to remark that this result becomes to be important,  because  we have  extended  the number of dynamical variables by considering  the 1-form conexion $A^{IJ}$ and  the two-form $R_{IJ}$ as independent. Nevertheless,  we have not lost  the symmetries of  Pontryagin theory. \\
We can  compare  the results reported in this paper with the reported in \cite{11b} and   \cite{11c} where the Hamiltonian analysis for topological theories has been performed on a smaller phase space context. However,  in our work we have identified the complete form of the first class  and second class constraints,  the extended Hamiltonian and the gauge transformations. In this sense,  our methodology  extends and complete the  previous ones, thus we are   showing a clear advantage    when is   applied  a pure Dirac  method for the theories under study. 
\newline
\newline
\newline
\noindent \textbf{VI. Conclusions and prospects}\\[1ex]
 \\
In this paper, we could present a clear and  consistent  application of a pure Dirac's method for constrained systems.  By working with the original phase  space we could perform the Hamiltonian dynamics  for the Chern-Simons theory and for the  Pontryagin invariant. With the present analysis we could identify for both  theories the extended action, the extended Hamiltonian and the full constraints program. The correct identification  of the constraints as first and second class,  allowed us carry out the counting of degrees of freedom, concluding that the theories under study corresponds to  topological field theories. We could observe,  that  Chern-Simons theory and  the Pontryagin  invariant has a closed relation at   Hamiltonian level. From one side,  the Chern-Simons theory has presence of  first and second class constraints. From other side,  Pontryagin theory presented only first class constraints and reducibility conditions among  them.  Thus, both theories are related by the action  (\ref{eq1a})  and his  Hamiltonian study  indicates that the theories  shares the principal symmetries namely;  zero degrees of freedom and diffeomorphisms as gauge transformations. \\
On the other  hand, by extending  the original configuration space  for the Pontryagin theory  we could perform the Hamiltonian analysis for this modified theory. We could observe that this extended theory shares the same symmetries with  unmodified  Pontryagin. Nevertheless,    the price to pay for extending the configuration space   is that now we have the presence of second class constraints while for unmodified Pontryagin we do not have it. But, by considering  the second class constraints as strong equations, we can reproduce the results found for the  Pontryagin invariant.\\
As final conclusion of this paper, the results presented in  this work  allowed us to understand at Hamiltonian level the existing relation  among Chern-Simons theory and the Pontryagin invariant. In this manner, we expect  that these results  will be  useful to develop  the  quantum treatment for both theories,  and thus,  to obtain a best  understanding for the quantum theory. In particular, the results of this article  presents  a best classical description than the results   reported  in \cite{11} and \cite{11b}. Therefore,  we can analyze the  quantization aspects for the Pontryagin theory by using the context presented  in this work,  taking on to account the  original configuration space.  However, this important part will be reported in forthcoming works.   
\newline
\newline
\newline
\noindent \textbf{Acknowledgements}\\[1ex]
This work was supported by CONACyT under grant 95560. Leopoldo Carbajal wants to thank to  IFUAP's  Theoretical Physics   group under grant promep: 169. 

\end{document}